\documentclass[fleqn,twoside]{article}
\usepackage{espcrc2}
\usepackage{url}

\usepackage{graphicx}

\title{A Measurement of the UHECR Spectrum with the HiRes FADC Detector}

\author{A.Zech
  \address[Rutgers]{Rutgers, 
    The State University of New Jersey  \\
    Department of Physics and Astronomy \\
    Piscataway, New Jersey, USA 08854 \\
    aszech@physics.rutgers.edu}
  presented on behalf of the High Resolution 
    Fly's Eye Collaboration}

\begin{document}

\begin{abstract}
We have measured the energy spectrum of ultra-high energy cosmic rays
(UHECR) with the HiRes FADC detector (HiRes-2) in monocular mode. A detailed
Monte Carlo simulation of the detector response to air showers has
been used to calculate the energy dependent acceptance of the air
fluorescence detector. The measured spectrum complements the
measurement by the HiRes-1 detector down to lower energies.
Systematic effects of the assumed input spectrum and composition 
on the aperture are presented, as well as systematics due to the
atmosphere. 
\end{abstract}

\maketitle

\section{Introduction: Mono vs. Stereo}

The two air fluorescence detectors of the HiRes experiment provide
stereoscopic observation of extensive air showers (EAS), initiated in the 
atmosphere by ultra-high energy cosmic rays. 
The stereo view leads to an improved resolution in the 
reconstructed arrival direction and energy of the primary cosmic ray
particle due to better constraints on the location of the shower axis.
 However, even if observation with both ''eyes'' is the 
preferred method, there are good reasons for the analysis of monocular
data. Since the HiRes-1 detector began operation two years before
HiRes-2, the HiRes-1 dataset provides larger statistics than the stereo
data. This is especially significant in the measurement of the cosmic
ray energy spectrum at energies around the GZK flux suppression.

The
HiRes-2 monocular dataset, which will be presented here, offers a different
advantage: It contains well reconstructed events at energies
lower than HiRes-1 mono and stereo events. The HiRes-2 detector uses
FADC electronics to record signals at a frequency of 10 MHz,
which leads to a better time resolution than is achieved with the
sample-and-hold electronics of the HiRes-1 detector. Furthermore, the 
two rings of mirrors of HiRes-2 provide twice the coverage in
elevation angle that is achieved with the single ring of mirrors of
the HiRes-1 detector. Events recorded with 
HiRes-2 can be reliably reconstructed down to energies of about $10^{17} eV$. 
This is also about a decade lower in energy than the
lower limit for stereo events, which is
constrained by the separation between the two detectors: Events with
the lowest observable energy lie half-way between the two detectors
and have thus a distance of about $6 \; km$ from each
detector, which sets a lower limit to their observable
energies.

 The HiRes-2 measurement of the UHECR spectrum
is intended to complement the HiRes-1 measurement at lower energies, 
where the features of the ''second knee'' and the ''ankle'' provide  
recognizable signatures of the cosmic ray spectrum, which can be used to 
compare the HiRes measurements to other experiments. 
        
\section{Monte Carlo Simulation}

\begin{figure}[htb]
\vspace{-1.5pc}
  \includegraphics[width=\columnwidth]{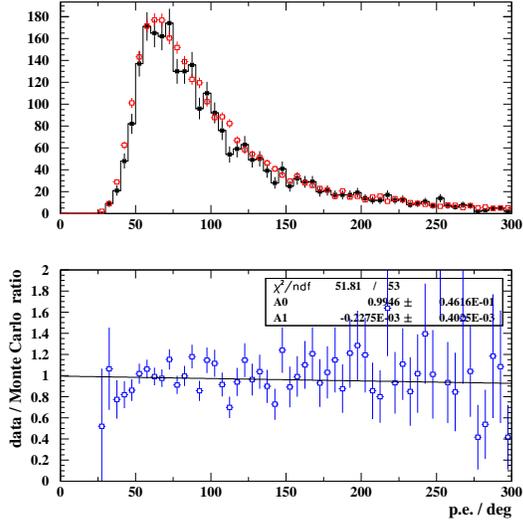}
\vspace{-2.6pc}
  \caption{{\bf top:} Distribution of photoelectrons per degree of track
           for data (filled squares) and MC (open squares). 
           Histograms have been normalized to cover the same area. 
           {\bf bottom:} Ratio of data over MC with linear fit. }
  \label{datamc}
\end{figure}

The calculation of the cosmic ray energy spectrum from the measured event
distribution is a problem of unfolding the true spectrum of cosmic
rays at their arrival in the earth's atmosphere from the distorsions
of the detector response, i.e. the acceptance of the detector and its 
limited resolution. 

The purpose of the Monte Carlo (MC) simulation in the HiRes experiment is
to give an accurate description of the geometry and energy dependent 
acceptance of the detector and of its resolution. 
The simulation consists
basically of two parts: an air shower generator and a detector
response MC. In the first part of the simulation, large sets of
air showers are generated with several discrete energies and 
with different
primary particles, using CORSIKA~\cite{corsika} and
QGSJet~\cite{qgsjet}. Their profiles are 
saved in a library of air showers. From this library, the individual 
profiles can be read into the detector response MC and used to simulate
EAS at different
geometries. Energies in the detector response MC are chosen from a
given continuous input spectrum. Profiles from library showers
generated at a nearby, discrete energy are scaled to the chosen energy. 

The detector response simulation includes:
generation of fluorescence and \v{C}erenkov light at the shower,
propagation of light through the atmosphere, ray tracing of photons
through the optical path of the detector, PMT response
to the signal, simulation of noise, electronics and trigger
simulation. Events accepted by the trigger are written out 
in the same format as the data.

Two databases have been generated and are read by the detector
response MC in order to allow event simulation under the
exact data taking conditions. A trigger database contains information
on the livetime of the detector, dead mirrors and variable trigger settings. An
atmospheric database provides hourly measurements of the aerosol
content of the atmosphere. An average atmosphere has been used in 
this analysis to allow a direct comparison with the HiRes-1
measurement.

\begin{figure}[htb]
\vspace{-1.5pc}
  \includegraphics[width=\columnwidth]{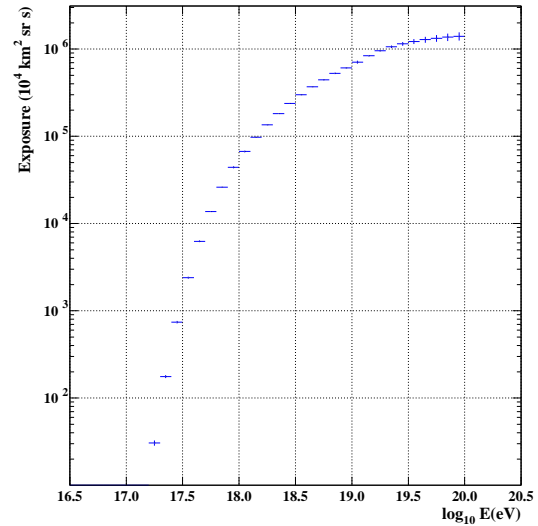}
\vspace{-2.6pc}
  \caption{Exposure as a function of energy, calculated from MC simulations.}
  \label{exposure}
\end{figure}

\section{Event Reconstruction and Analysis}

As the air shower develops in the atmosphere, its image sweeps across
the photomultiplier tube clusters, causing a track of triggered
phototubes. The first step in the reconstruction of an event is the 
determination of the shower-detector plane from a fit to this track.
The geometry of the shower axis within this plane and the distance
of the shower from the detector are found in the HiRes-2 monocular 
reconstruction with a fit of the times of triggered phototubes
versus their angle along the shower track. Once the location and 
geometry of the shower axis have been determined, the charged particle 
profile of the air shower is reconstructed from the distribution of
recorded FADC counts in triggered phototubes. Absolute calibration
of the photomultiplier tubes with a calibrated light source and
measurements of the aerosol density of the atmosphere with steerable
lasers ensure the accurate calculation of the shower profile from 
the FADC signals. Using the atmosphere as a calorimeter, we can calculate the
total deposited energy of the shower from the charged particle
profile after subtraction of \v{C}erenkov light. Adding about $10 \%$ 
for ''missing energy'', which is deposited into the ground in muons
and neutrinos or lost in nuclear excitations, yields the total energy
of the primary cosmic ray particle.

The same quality cuts and the same reconstruction procedure are
applied to both data and simulated
events, which allows us to directly compare all the details of our
MC simulation against the actual experiment. An extensive set of
data-MC comparison plots---including e.g.\ the distance of
air showers, their angular distributions, number of triggered mirrors
and phototubes, reconstructed profiles and energies, etc.---tests
how well our MC simulates the actual detector. An example for the 
overall very good agreement between data and MC is shown in 
Figure~\ref{datamc}: The distributions of photoelectrons per
degree of track are measures of the amount of light that is seen
in the data and generated in the MC. The simulation closely follows the data
distribution. More details on the HiRes-2 analysis can be found 
in~\cite{longpaper}.

With our MC simulation proven to be realistic, we can calculate the
acceptance of the HiRes-2 detector, which is given by the ratio of 
fully reconstructed events ($R_{MC}$) over all generated events ($G_{MC}$) 
in each energy
bin. Multiplying the acceptance by the geometrical aperture $A \Omega$ and the
livetime $t$ of the detector (about 531 hours of analyzed data) yields
the exposure of the detector. We fit 
the simulated exposure to 
an appropriately chosen function in order to smooth out statistical
fluctuations. The result is shown in Figure~\ref{exposure}.

The differential flux $J$ in each energy bin can now be calculated as:
\begin{displaymath}
J(E_i)= N(E_i)*\frac{1}{\Delta E}*\frac{1}{\frac{R_{MC}(E_i)}{G_{MC}(E_i)}*A
  \Omega*t} 
\label{espec}
\end{displaymath}
where $N(E_i)$ is the number of data events in the energy bin and
$\Delta E$ is the binwidth.

\begin{figure}[htb]
\vspace{-1.5pc}
  \includegraphics[width=\columnwidth]{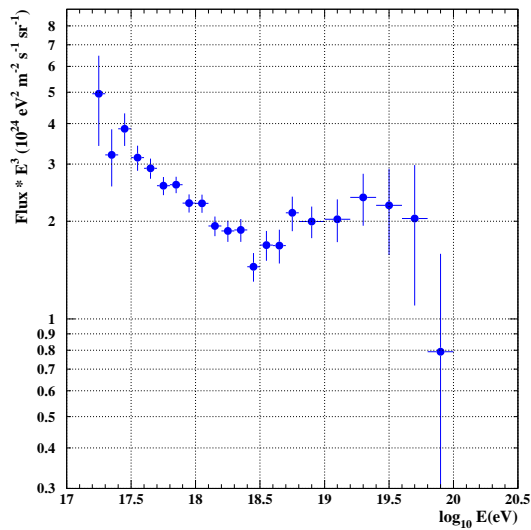}
\vspace{-2.6pc}
  \caption{Energy spectrum ($J*E^3$) measured by the HiRes-2 detector
           in monocular mode.}
  \label{hr2_spec}
\end{figure}

\section{The HiRes-2 Spectrum}

\begin{figure}[htb]
\vspace{-1.5pc}
  \includegraphics[width=\columnwidth]{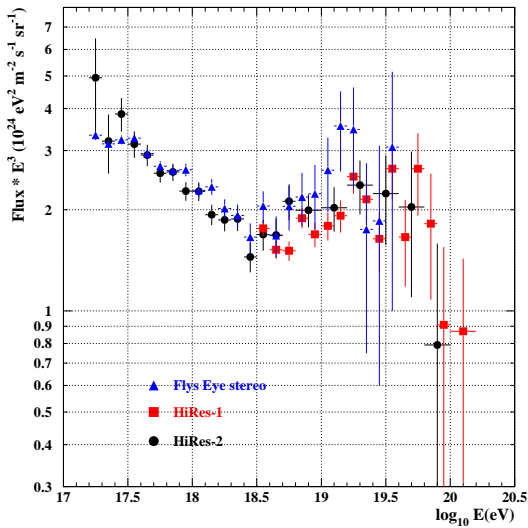}
\vspace{-2.6pc}
  \caption{Energy spectra ($J*E^3$) measured by HiRes-1 (squares),
           HiRes-2 (circles) and Fly's Eye stereo (triangles).}
  \label{hr12fe_spec}
\end{figure}

The energy spectrum measured by the HiRes-2 detector on clear, 
moonless nights from December 1999 to September 2001 ($\sim 2700$
events) is shown in Figure~\ref{hr2_spec}. We observe the ''ankle''
at about $10^{18.5} eV$. Our updated measurement is consistent with
the GZK flux suppression, but statistics at the high energy end
are low.
Figure~\ref{hr12fe_spec} shows the spectra measured by HiRes-1, HiRes-2 and
the Fly's Eye stereo experiment superimposed. The two monocular HiRes
spectra agree with each other and with our predecessor
experiment. There is a slight difference in the slopes of the HiRes 
and Fly's Eye spectra above the ''ankle'', which is most likely due to a 
different atmospheric model used in the Fly's Eye experiment. The low
energy end of the HiRes-2 spectrum is consistent with the Fly's Eye
spectrum. However, due to low statistics, HiRes cannot claim detection of 
the ''second knee'', seen by Fly's Eye at $10^{17.6} eV$.

\vspace{-1.5pc}

\section{Studies of Systematics}
The main systematic uncertainties in the HiRes monocular spectra have
been reported in~\cite{prl}: Uncertainties in the absolute phototube
calibration, fluorescence yield, ''missing energy'' correction and 
atmospheric calibration add up to a total uncertainty in the flux of
$ \pm 31 \%$. Here, we want to discuss a potential impact on the 
calculated acceptance from our assumptions about the MC input energy
spectrum and input composition, and the assumption of an average
atmosphere in our simulation and analysis.

\subsection{Input Spectrum}
\begin{figure}[htb]
\vspace{-1.5pc}
  \includegraphics[width=\columnwidth]{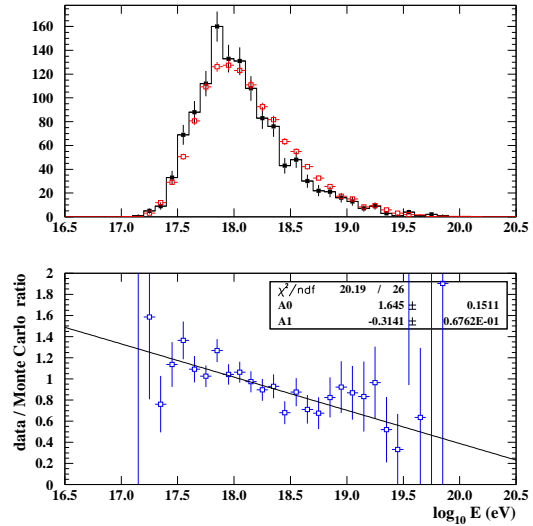}
\vspace{-2.6pc}
  \caption{{\bf top:} Energy distribution of real events (filled squares)
                and simulated events (open squares) using an $E^{-3}$ input
                spectrum. 
           {\bf bottom:} Ratio of data over MC.}
  \label{e3inp}
\end{figure}

Using MC simulations to determine the acceptance of the detector 
could potentially introduce a bias into the analysis due to a wrong 
model used in the MC. This potential bias can be
estimated by calculating the acceptance for different model
assumptions in the MC~\cite{cowen}. We have varied the shape of 
the input spectrum,
from which MC event energies are chosen. Figure~\ref{e3inp} shows a 
plot comparing the energy distributions of a MC set 
using a simple $E^{-3}$ power law as input with about half of our analyzed
data. The disagreement of the 
data with this assumption can be seen clearly from the tilted ratio
plot.  

\begin{figure}[htb]
\vspace{-1.5pc}
  \includegraphics[width=\columnwidth]{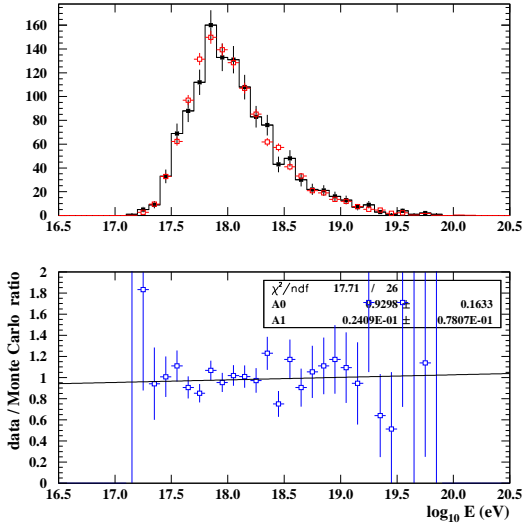}
\vspace{-2.6pc}
  \caption{{\bf top:} Energy distribution of real events (filled squares)
                and simulated events using a realistic input spectrum
                as described in the text (open squares). 
           {\bf bottom:} Ratio of data over MC.}
  \label{ourinp}
\end{figure}

Our regular MC uses an input spectrum whose shape fits the Fly's Eye
stereo spectrum below the ''ankle'' and follows a straight line fit
to the HiRes-1 data for higher energies. The GZK feature has not been 
included. A comparison plot for this input spectrum
is shown in Figure~\ref{ourinp}. Here, the agreement is very good.

We have calculated acceptances for both MC sets. The bias we are
avoiding by using a realistic input spectrum is of the order of
$\pm 20 \%$. This bias would translate directly into the measured
spectrum, if one assumed a wrong (i.e. $E^{-3}$) input spectrum
in the MC simulation. 

\subsection{Input Composition}
We determine the fraction of showers initiated by light and heavy,
i.e.\ proton and iron, cosmic rays in the MC from composition measurements 
by the HiRes/MIA~\cite{HiResMIA} and HiRes stereo~\cite{HiResStereo} 
experiments. 

The uncertainties in the HiRes/MIA measurement that translate into our
spectrum calculation add up to $\sim 5 \%$. Their sources are the
detector calibration, the aerosol component of 
the atmosphere and the statistical uncertainty of a fit to the
HiRes/MIA data. A $\sim 10 \%$ uncertainty in the fluorescence 
yield is common to both HiRes and HiRes/MIA and was therefore not
included. The difference in the predictions of pure iron and pure
proton maximum shower depths ($X_{max}$) between different hadronic 
interaction models was not taken into account either, since we are not
concerned about the fraction of real proton and iron showers here, but
only about the fraction of showers with a certain $X_{max}$.

\begin{figure}[htb]
\vspace{-1.5pc}
  \includegraphics[width=\columnwidth]{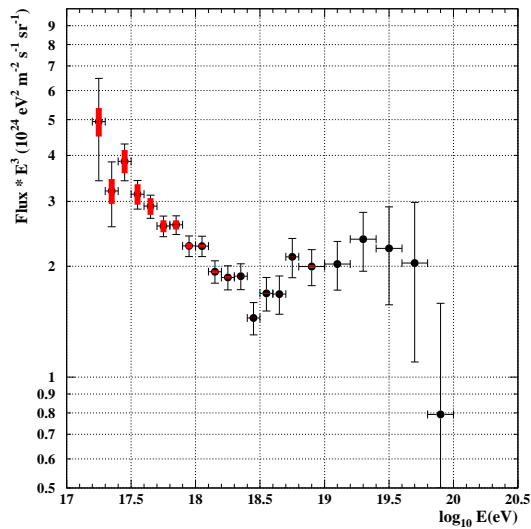}
\vspace{-2.6pc}
  \caption{HiRes-2 energy spectrum with systematic uncertainties
    (thick error bars) corresponding to a $\pm 5 \%$ change in the
    proton fraction of the MC.}
  \label{compsys}
\end{figure}

By generating two MC sets with pure proton and pure iron showers,
we can calculate the effect a $\pm 5 \%$ change in the proton fraction
would have on the
acceptance. The uncertainty in the final spectrum from such a
variation is shown in Figure~\ref{compsys}. At the low energy end of
the spectrum, the acceptance for iron cosmic rays is lower because
iron showers develop higher up in the atmosphere and are more likely
to be outside of our elevation coverage than proton showers. This leads
to larger uncertainties at lower energies. For energies above $10^{18}
eV$, where we use the HiRes stereo composition measurement, no
difference was seen in the acceptances for iron and proton showers.

\subsection{Atmospheric Database}
Our current analysis uses a measurement of the average aerosol 
content of the atmosphere. We have studied the effect on the energy
resolution of using an atmospheric database with hourly entries
instead of the average. A MC set has been generated with use of the
database to simulate data from September 2000 to March 2001. Our
database covers $80 \%$ of all nights in this period. The MC set
has been reconstructed first with the average and then with the database.
A Gaussian fit to the energy resolution has a
$\sigma$ of $16.2\%$ for the reconstruction using the database and
$17.5\%$ when using the average. 

\begin{figure}[htb]
\vspace{-1.5pc}
  \includegraphics[width=\columnwidth]{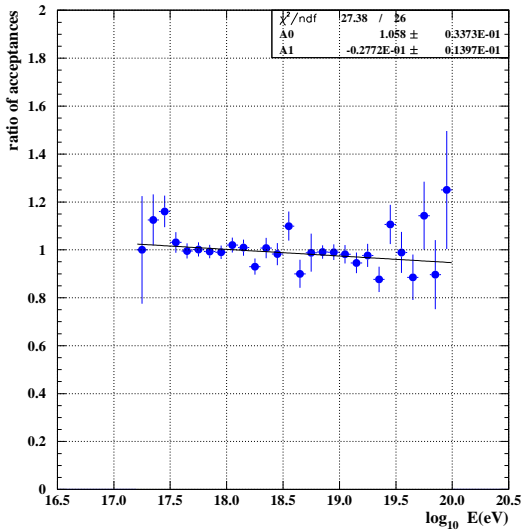}
\vspace{-2.6pc}
  \caption{Ratio of acceptances for a MC set generated using an
           atmospheric database and reconstructed with the
           database (numerator) and an average atmosphere (denominator).}
  \label{apratio_adb}
\end{figure}

An acceptance has been calculated in both cases. The ratio of
acceptances can be seen in Figure~\ref{apratio_adb}: The acceptance
for the MC set reconstructed with database is in the numerator, the
acceptance for the same MC set, but reconstructed with the average,
in the denominator. Using the database in the reconstruction does not
have a significant impact on the calculated aperture. 

Data from September 2000 to March 2001 have been analyzed with the average
and with the atmospheric database. We have not found any systematic
shift in the reconstructed energies. When we examined the highest
energy events ($>10^{19.5} eV$), we found that using the database 
shifts one
event from the bin centered at $10^{19.55} eV$ to the bin centered at
$10^{19.45} eV$. This leads to a reduction of the events above
$10^{19.5}$ from 7 to 6 in this dataset.

\section{Conclusions}
\begin{itemize}
\item Our measurement of the UHECR spectrum with the HiRes FADC
  detector in monocular mode is consistent with the GZK flux 
  suppression at $10^{19.8} eV$ and with the Fly's Eye stereo
  spectrum. The ''ankle'' is observed at an energy of about $10^{18.5} eV$. 
\item Detailed Monte Carlo simulation is used to calculate the 
  acceptance of the detector, after being tested with data-MC comparisons.
\item We avoid a bias of about $\pm 20 \%$ in our spectrum by  
  using a realistic assumption about the input energy spectrum in the
  MC.
\item Systematic uncertainties caused by uncertainties of the input
  composition in the MC do not exceed our statistical errors. 
\item We have tested the impact of using an atmospheric database
  with hourly entries rather than an average atmosphere in the
  reconstruction for about half of our analyzed data. We have not
found a significant difference. 

\end{itemize}


\begin{thebibliography}{9}
  
\bibitem{corsika} D.~Heck, J.~Knapp, J.N.~Capdevielle, G.~Schatz and
  T.~Thouw, Report FZKA 6019 (1998), Forschungszentrum Karlsruhe.
  
\bibitem{qgsjet} N.N.~Kalmykov, S.S.~Ostapchenko and A.I.~Pavlov,
  Nucl.  Phys. B (Proc. Suppl.) {\bf 52B}, 17, (1997).

\bibitem{longpaper} R.U.~Abassi {\it et al.}, submitted to
  Astropart. Phys., arXiv: astro-ph/0208301

\bibitem{prl} R.U.~Abassi {\it et al.}, Phys.~Rev.~Lett. {\bf 92},
  (2004) 151101., arXiv: astro-ph/0208243

\bibitem{cowen} G. Cowen, Statistical Data Analysis, Oxford Science
   Publ. (1998)

\bibitem{HiResMIA} T.~Abu-Zayyad, {\it et al.}, Phys.~Rev.~Lett. {\bf
    84} (2000) 4276.
  
\bibitem{HiResStereo} G.~Archbold and P.V.~Sokolsky (for the HiRes
  Collaboration), Proc. of the 28th Int. Cosmic Ray Conf. (2003) 405.

\end{thebibliography}
\end{document}